\documentclass[aps,prd,preprint,superscriptaddress,tightenlines,%
nofootinbib]{revtex4}
\newcommand{\PRE}[1]{{#1}}   

\usepackage{bm}
\usepackage{epsfig}

\newcommand{\mathbb}[1]{\mbox{\Bbb #1}}

\def\lbldef#1#2{\expandafter\gdef\csname #1\endcsname {#2}}

\def\href#1#2{#2}

\newcommand{\ber}{\begin{eqnarray}}
\newcommand{\eer}{\end{eqnarray}}

\newcommand{\beqar}{\begin{eqnarray}}

\newcommand{\eeqar}{\end{eqnarray}}

\newcommand{\dsl}
  {\kern.06em\hbox{\raise.15ex\hbox{$/$}\kern-.56em\hbox{$\partial$}}}

\newcommand{\eeqarr}{\end{eqnarray}}
\newcommand{\ZZ}{{\rm \kern 0.275em Z \kern -0.92em Z}\;}

\begin{document}

\preprint{
\hfil
\begin{minipage}[t]{3in}
\begin{flushright}
\vspace*{.4in}
NUB--3250--Th--04\\
MIT--CTP--3524\\
hep-ph/0408284
\end{flushright}
\end{minipage}
}

\title{
\PRE{\vspace*{1.5in}} Probing Split Supersymmetry with Cosmic Rays
\PRE{\vspace*{0.3in}} }

\author{Luis Anchordoqui}
\affiliation{Department of Physics,\\
Northeastern University, Boston, MA 02115
\PRE{\vspace*{.1in}}
}

\author{Haim Goldberg}
\affiliation{Department of Physics,\\
Northeastern University, Boston, MA 02115
\PRE{\vspace*{.1in}}
}

\author{Carlos Nu\~nez}
\affiliation{Center for Theoretical Physics,\\
Massachusetts Institute of Technology, Cambridge, MA 02139
\PRE{\vspace*{.5in}} }

\date{August 2004}
\PRE{\vspace*{.5in}}
\begin{abstract}

\noindent A striking aspect of the recently proposed split
supersymmetry is the existence of heavy gluinos which are
metastable because of the very heavy squarks which mediate their
decay. In this paper we correlate the expected flux of these
particles with the accompanying neutrino flux produced in
inelastic $pp$ collisions in distant astrophysical sources.  We
show that an event rate at the Pierre Auger Observatory of
approximately 1 yr$^{-1}$ for gluino masses of about 500~GeV is
consistent with existing limits on neutrino fluxes. Such an 
event rate requires powerful cosmic ray engines able to 
accelerate particles up to extreme energies, somewhat above 
$5 \times 10^{13}$~GeV. The extremely
low inelasticity of the gluino-containing hadrons in their
collisions with the air molecules makes possible a distinct
characterization of the showers induced in the atmosphere. Should
such anomalous events be observed, we show that their cosmogenic
origin, in concert with the requirement that they reach the Earth
before decay, leads to a lower bound on their proper lifetime of
the order of 100 years, and consequently, to a lower bound on the
scale of supersymmetry breaking, $\Lambda_{\rm SUSY} >2.6 \times
10^{11}$~GeV. Obtaining such a bound is not possible in collider
experiments.

\begin{center}
\end{center}
\end{abstract}

\maketitle

\section{General idea}

The standard model (SM) of particle physics has had
outstanding success in describing all physical phenomena up to
energies $\sim 500$~GeV~\cite{Hagiwara:pw}. Nonetheless, there is a
general consensus on that it is not a fundamental theory of
nature (apart from the fact that the SM does not include gravity):
with no new physics between the energy
scale of electroweak unification ($M_{W} \sim 10^2$~GeV) and
the vicinity of the Planck mass ($M_{\rm Pl} \sim 10^{19}$~GeV)
the higgs mass must be fine-tuned to an accuracy of order
$(M_W/M_{\rm Pl})^2$ to accommodate this enormous desert. The leading
contender for the elaboration of the desert  has been the
supersymmetric extension of the SM~\cite{Dimopoulos:1981zb}.
Supersymmetry (SUSY) posits a ``complete democracy'' between
integral and half-integral spins, implying
the existence of many as-yet-undiscovered superpartners. Thus, if SUSY
can serve as a theory of low energy interactions, it must be a
broken symmetry. The most common assumption is that the minimal low energy
effective supersymmetric theory (MSSM) has a breaking scale of
order $\Lambda_{\rm SUSY} \sim 1$~TeV, thus avoiding 't~Hooft
naturalness problem with the higgs mass.

MSSM has  a concrete advantage in embedding the SM in a grand
unified theory: the supersymmetric beta functions for
extrapolating the measured strengths of the strong,
electromagnetic, and weak couplings lead to  convergence at a
unified energy value of the order $M_{\rm  GUT }\sim
10^{16}$~GeV~\cite{Dimopoulos:1981yj}. The model, however, is not
free of problems. In particular, dimension four $R$-parity
violating couplings in the superpotential  yield unacceptably
large proton decay rates and neutrino masses. This can be readily
solved by imposing  $R$-parity conservation, which as a byproduct
ensures the stability of the lightest SUSY particle, making it a
possible candidate for cold dark matter~\cite{Goldberg:1983nd}. However,
there are other problems in the MSSM: dimension five operators tending
to generate excessive proton
decay; new  CP-violating phases which require suppression for
agreement with limits on electric dipole moment limits; and
excessive  flavor violations, due {\em e.g.,} to the absence of a
complete flavor degeneracy in the K\"ahler potential of minimal
supergravity~\cite{Ibanez:1992hc}.

Of course, the fine-tuning involved in accommodating the above constraints is
miniscule in comparison to that required in generating  a
cosmological constant that satisfies 't~Hooft naturalness. Recent experimental
data~\cite{Bahcall:1999xn}
strongly indicate that the universe
is expanding in an accelerating phase, with an  effective de Sitter constant $H$
that nearly saturates the upper bound given by the present-day value of the
Hubble constant, i.e., $H \alt H_0 \sim 10^{-33}$~eV. According to the Einstein field equations,
$H^2$ provides a measure of
the scalar curvature of the space and is related to the vacuum energy density,
$\epsilon_4,$ according to
$M_{\rm Pl}^2 H^2 \sim \epsilon_4.$ However, the ``natural'' value of $\epsilon_4$ coming from the
zero-point energies of known elementary
particles is found to be at least $\epsilon_4 \sim \Lambda_{\rm SUSY}^4,$
yielding $H \agt 10^{-3}$~eV.
The failure of 't~Hooft naturalness then centers on the following question: why is the vacuum energy
determined by the Einstein field equations 60 orders of magnitude smaller than
any ``natural'' cut-off scale in
effective field theory of particle interactions, but not zero? Nowadays,
the only existing framework which
can address  aspects of this question is the anthropic approach~\cite{Weinberg:dv}.

Very recently Arkani--Hamed and Dimopoulos (AD) have looked at SUSY from a different
angle~\cite{Arkani-Hamed:2004fb}.  In their
model, the scale of SUSY breaking is pushed to a very high energy
(say, $\Lambda_{\rm SUSY} \sim 10^{13}$~GeV) and a  higgs
mass of order TeV is recovered by
invoking fine tuning.  For this breaking scale,
the bosonic superpartners are heavy,
while the extra fermions retain TeV-scale
masses thanks to protection by chiral symmetry. (We follow Giudice and
Romanino~\cite{Giudice:2004tc} in adopting the designation
``split SUSY''  for the AD model). This scenario preserves
the achievements of the MSSM
while resolving the problems mentioned above. In particular, analyses of
one loop~\cite{Arvanitaki:2004eu} and
two loops~\cite{Giudice:2004tc} running of the RG equations,
show that the AD scenario preserves
unification of couplings. Moreover,  aside from the light higgs tuning,
the other flaws inherent to the MSSM elegantly disappear when the scalar superpartners decouple.

The AD model can be discussed in the same  anthropic framework adopted for examination of
the cosmological constant. Recent
investigations in String Theory have applied a statistical approach  to the
enormous ``landscape'' of
vacua present in the theory~\cite{Susskind:2003kw}. Among this vast number of
metastable vacua, there can be
small subset ${\cal O} (10^{40})$ exhibiting low scale SUSY breaking, a TeV-scale
higgs, as well as the
remaining traditional MSSM physics~\cite{Douglas:2003um}.  However, the fine
tuning required to achieve
a small cosmological constant implies the need of a huge number of vacua, far more than
the ${\cal O} (10^{40})$
characterizing low-scale
SUSY breaking~\cite{Dine:2004is}. Remarkably, if one
posits high-scale SUSY breaking and superpartners widely separated in
mass, ${\cal O} (10^{200})$ vacua become available, enough to fine
tune both the cosmological constant and the higgs mass.

It is therefore instructive to explore how drastically the AD scenario can change the
phenomenology of conventional MSSM. Prospects for probing split SUSY at the LHC~\cite{Zhu:2004ei}
as well as in dark matter searches~\cite{Pierce:2004mk} have been recently developed.
In what follows
we show that  cosmic ray data may also provide important information about the AD scenario.
As a principal result of this paper, we will delineate
conditions under which one can set a {\it lower} bound on the SUSY breaking scale in a region
of parameter space far beyond that  probed at the LHC.

An intriguing prediction in this scenario, which represents a radical departure
from the MSSM, is the longevity of the gluino~\cite{Demir:2004kf}.
As mentioned above,
in split SUSY
the squarks are
very massive and so gluino  decay via virtual squarks becomes strongly
suppressed, yielding a
$\tilde g$ lifetime of the order of~\cite{Arkani-Hamed:2004fb}
\begin{equation}
\tau_0 \simeq  \frac{64 \pi^3 \Lambda_{\rm SUSY}^4}{M_{\tilde g}^5}
 \approx 10^{7} \left(\frac{{\rm TeV}}{M_{\tilde g}}\right)^5
\left(\frac{\Lambda_{\rm SUSY}}{10^{13}~{\rm GeV}}\right)^4~{\rm yr}
\label{tau0}
\end{equation}
where $M_{\tilde g}$ is the gluino mass. Very strong limits on
heavy isotope abundance in turn require the gluino to decay on
Gyr time scales, leading to an upper bound for the scale of SUSY
breaking ${\cal O} (10^{13})$~GeV~\cite{Arkani-Hamed:2004fb}.
Because of the large mass of the gluino, the threshold for
inelastic scattering on the cosmic microwave background is $\agt
10^{14}$ GeV~\cite{Farrar:1996rg}, allowing ultra-high energy
gluino-containing hadrons (``$G$'s'') to reach us unimpeded from
cosmological distances.

In this work we study the possibility of $G$-detection with
cosmic ray observatories. To this end, in  Sec.~\ref{showers} we
discuss the main characteristics of cascades induced by
$G$-hadrons and estimate the sensitivity of the Pierre Auger Observatory (PAO).
After that, in Sec.~\ref{Gproduction},  we
correlate the expected flux of $G$-hadrons with the accompanying neutrino
flux produced in inelastic $pp$ collisions in distant astrophysical sources.
We show that an event rate $\approx  1$~yr$^{-1}$ at PAO for gluino masses
of about 500~GeV is consistent with existing limits on neutrino fluxes. The actual
observation of  a few $G$-events will then
directly imply a lower bound on $\Lambda_{\rm SUSY}$. The details
of this interesting possibility are presented in
Sec.~\ref{lambdaS}. Section~\ref{conclusions} contains our
conclusions.

\section{Characteristics of air showers initiated by $\bm{G}$--hadrons}
\label{showers}

The interaction of a high energy cosmic ray in the upper
atmosphere gives rise to a roughly conical cascade of particles
that reaches the Earth's surface in the form of a giant
``saucer'', traveling at nearly the speed of light. In the case of
proton-- or nucleus--induced cascades, the leading particle and
other high energy hadrons (mostly pions) in the shower core
readily cascade to lower energies as they interact with the air
molecules. Because of the prompt decay of neutral pions, 1/3 of
the energy in each of these interactions transits into energetic
$\gamma$--rays. Electromagnetic subshowers are then initiated: the
high energy photons produce pairs that lose energy by
bremsstrahlung and ionization before annihilation into a new
photon at lower energy. Eventually, the average energy per
particle drops below a critical energy $\epsilon_0 \sim 86$~MeV at
which point ionization takes over from bremsstrahlung and pair
production as the dominant energy loss mechanism. The changeover
from radiation losses to ionization losses depopulates the shower,
and defines $X_{\rm max},$ the longitudinal coordinate of maximum
multiplicity.

The number of muons (and neutrinos) does not increase linearly
with energy, because at higher energy more generations are
required to cool the pions to the point where they are likely to
decay before interaction. Production of extra generations results
in a larger fraction of the energy being lost to the
electromagnetic cascade, and hence a smaller fraction of the
original energy being delivered to the $\pi^{\pm}$. The electrons,
positrons and photons are thus the most prolific constituents in
the thin disk of particles showering towards the ground, and most
of the energy  (about 90\%)  is dissipated in the electromagnetic
cascade.

By the time they reach the ground, relatively vertical showers
have evolved fronts with a curvature radius of a few km, and far
from the shower core their constituent particles are well spread
over time, typically of the order of a few microseconds.  For such
a shower both the muon component and a large portion of the
electromagnetic component survive to reach the ground. For
inclined showers the electromagnetic component is absorbed long
before reaching the ground, as it has passed through the
equivalent of several vertical atmospheres: 2 at a zenith angle
$\theta = 60^\circ$, 3 at $70^\circ$, and 6 at $80^\circ$. In
these showers, only high energy muons created in the first few
generations of particles survive past 2 equivalent vertical
atmospheres. The rate of energy attenuation for muons is much
smaller than it is for electrons, thus the shape of the resulting
shower front is  very flat (with curvature radius above $100$~km),
and its time extension is very short (less than
$50$~ns)~\cite{Anchordoqui:2004xb}.

The energy lost by a $G$ during collision with nucleons is
primarily through hard scattering~\cite{Berezinsky:2001fy}. This
implies a  fractional energy loss per collision, $K_{\rm
inel} \approx 1~{\rm GeV}/M_G$. A heuristic justification for this
result is as follows: consider the inclusive process $G N
\rightarrow G X.$ It is a simple kinematic exercise to show that
the minimum momentum transfer for $M_X^2, M_G^2 \ll s$ is given by
$|t_{min}|\simeq M_G^2\ M_X^4/s^2,$ where $\sqrt{s}$ is the
center-of-mass energy of the $G N$ collision. Kinematics also determine the
fractional energy loss, $K_{\rm inel} = M_X^2/s.$ Combining these
two results we obtain that $K_{\rm inel} \simeq |t_{min}|^{1/2}/M_G.$ The
hard scattering restriction in QCD requires $|t_{min}|\agt $ 1~GeV$^2$,
so that for a fiducial $M_G = 500$~GeV, we obtain $K_{\rm
inel} \approx 0.002.$ Note that, for $M_G=50$ GeV, our formula
agrees with the result in~\cite{Berezinsky:2001fy}.

In the case of a $G$-induced shower, the very low inelasticity of
$G$-air interactions implies the leading particle retains most of
its energy all the way to the ground, while the secondary
particles promptly cascade to low energies as for any other air
shower.  This results in an ensemble of mini-showers strung along
the trajectory of the leading particle.  Since the typical
distance between mini-showers is about 10 times smaller than the
extent of a single longitudinal profile, it is not possible to
resolve the individual showers experimentally. Instead one
observes a smooth envelope encompassing all the mini-showers,
which extends from the first interaction all the way to the
ground.  Monte Carlo simulations have been performed which confirm
this phenomenological description (see Fig.~3 in
Ref.~\cite{Berezinsky:2001fy}). The $G$--showers indeed present a
distinct profile: (1) there is  only a few percent probability for
its $X_{\rm max}$ to be mistaken for that of  a proton
shower~\cite{Albuquerque:1998va} (2) the flatness of the
longitudinal development is unique to the extremely low
inelasticity of the scattering, and can be easily isolated from
background.

The characteristics of $G$-induced showers observed at the ground
should also be distinct from those characteristic of  proton and
nucleus induced showers.  Each mini-shower generates a bundle of
muons which survive to the ground.  Since each bundle is produced
at a different slant-depth, the muon component of the shower front
exhibits a much more pronounced curvature than what one would
expect for a shower of the same energy interacting only in the
upper atmosphere.  The difference between a proton/nucleus and a
$G$ would thus be much more evident in the case of inclined
showers for which there is much less electromagnetic
contamination.   Specifically, a  highly inclined $G$-induced
shower produces many muon bundles, and so should exhibit a much
more curved shower front than a proton/nucleus shower with the
same energy and zenith angle.

The experimentally interesting region to search for $G$ events
then lies between $70^\circ < \theta < 90^\circ.$ The reduction of
the solid angle acceptance to larger than $70^\circ$ eliminates
the hadronic background. Moreover, there needs to be sufficient
pathlength for the $G$, with its low inelasticity, to lose
sufficient energy. The mean free path for ultra-high energy  $G$'s
is about 30~g cm$^{-2}$~\cite{Berezinsky:2001fy}, and the pathlength of the atmosphere at
$70^\circ$ is about 100 times this~\cite{Anchordoqui:2004xb}, allowing more than 20\% of the
energy to evolve in the air shower.

With this in mind, the $G$-event rate for a given cosmic ray
experiment is found to be
\begin{equation}
\frac{d {\cal N}}{dt} = \int_{E_{G,\,{\rm min}}}^{E_{G,\,{\rm max}}}
J_G(E_G)\, A(E_G) \, dE_G\, , \label{eventrate}
\end{equation}
where $A(E_G)$ is the hadronic aperture for $\theta > 70^\circ.$

There are two major techniques which can be employed in detecting
cosmic ray air showers. The most commonly used detection method
involves sampling the shower front at a given altitude using an
array of sensors spread over a large area.  Sensors, such as
plastic scintillators or \v{C}erenkov detectors are used to infer
the particle density and the relative arrival times of the shower
front at different locations. The muon content is usually sought
either by exploiting the signal timing in the surface sensors or
by employing dedicated detectors which are shielded from the
electromagnetic shower component. Inferring the primary energy
from energy deposits at the ground is not completely
straightforward, and involves proper modeling of both the detector
response and the physics of the first few cascade generations.
Another highly successful air shower detection method involves
measurement of the longitudinal development of the cascade by
sensing the fluorescence light produced via interactions of the
charged particles in the atmosphere. Excited nitrogen molecules
fluoresce producing radiation in the 300 - 400 nm ultraviolet
range, to which the atmosphere is quite transparent.  The shower
development appears as a rapidly moving spot of light whose
angular motion depends on both the distance and the orientation of
the shower axis. The integrated light signal is proportional to
the total energy deposited in the atmosphere. Fluorescence
observations can only be made on clear moonless nights, yielding a
duty cycle of about 10\%.

Over the next few years, the best observations of the extreme end
of the cosmic ray spectrum will be made by the PAO~\cite{Abraham},
which is actually comprised of
two sub-observatories. The Southern site is currently operational
and in the process of growing to its final size of $S \simeq
3000$~km$^2$. Another site is planned for the Northern hemisphere.
The PAO works in a hybrid mode, and when complete, each site will
contain 24 fluorescence detectors overlooking a ground array of
1600 water Cherenkov detectors. During clear, dark nights,  events
are simultaneously observed by fluorescence light and particle
detectors, allowing powerful reconstruction and cross-calibration
techniques. Simultaneous observations of showers using two
distinct detector methods will also help to control the systematic
errors that have plagued cosmic ray experiments to date.

For showers at inclination $\theta > 70^\circ$  and  energy
$E_{\rm sh}>10^9$~GeV the probability of detecting an event
falling within the physical area is roughly 1. Note that, from the
previous discussion,  $E_G\agt 5\ E_{\rm sh} \simeq 5\times 10^9$
GeV. The total aperture (2 sites) of the surface array is found to
be
\begin{equation}
A_{\rm PAO} = \int_{70^\circ}^{90^\circ} S\,\, \cos \theta
\,\,d\Omega\,\, \approx 2200 \,\, {\rm km}^2\,{\rm sr}\,\,,
\label{accept}
\end{equation}
where $S\,\cos \theta$ is the projected surface of the array in
the shower plane, and $d\Omega$ is the acceptance solid angle. From 
Eqs.~(\ref{eventrate}) and (\ref{accept}), one notes that an
event rate of $\approx 1$ yr$^{-1}$ requires an integrated flux
\begin{equation}
\int_{E_{G,\, {\rm min}}}^{E_{G,\,{\rm max}}} J_G(E_G)\,  dE_G\
\approx 1.4 \times 10^{-21} \,\, {\rm cm}^{-2} \,\,{\rm
s}^{-1}\,\, {\rm sr}^{-1}\,\,. \label{intflux}
\end{equation}
Requiring that events also trigger the fluorescence detectors
increases this flux by a factor of 10.

Detailed characteristics of ultra-high energy cosmic ray sources are largely
unknown~\cite{Anchordoqui:2002hs}, rendering a direct calculation of the expected $G$-flux speculative.
In the next section we assess the viability of this flux by comparing with existing
limits on gamma ray and neutrino fluxes.

\section{Production of $\bm{G}$--hadrons in astrophysical sources}
\label{Gproduction}

Among the non-thermal sources in the universe, radio--loud active galactic nuclei (AGNs) seem to be the
most important energetically. There are other interesting powerful sources, like gamma ray bursts and
cluster of galaxies; however, the non-thermal energy release in these astrophysical processes does not
come close that of AGNs. At radio frequencies, where very large baseline interferometers
can resolve the emission regions at milliarcsecond scale, many of radio--loud AGNs exhibit compact jets
of relativistic plasma which are remarkably well collimated, with opening angles about a few degrees or
less. The AGNs come in various disguises according to the orientation of their radio jets axes and
characteristics of the circum-nuclear matter in their host galaxies. The most extreme versions are
Fanaroff Riley radio-galaxies with the radio jet axes almost in the plane of the sky and  blazars
with the radio jet axes pointing close to the line of sight to the observer, yielding a significant flux
enhancement because of Doppler boosting.

A total of 66 blazars have been detected to date as GeV $\gamma$-ray sources by
the Energetic Gamma Ray Experiment Telescope (EGRET) on board of the Compton Gamma Ray Observatory
(CGRO)~\cite{Hartman:fc}. In addition, observations from ground-based \v{C}erenkov telescopes indicated
that at least in
2 of these sources the $\gamma$-ray spectrum can be traced to more than a TeV observed photon
energy~\cite{Punch:xw}.
The non-thermal emission of these powerful objects indicates a double-peak structure in the overall
spectral energy distribution. The first component (from radio to $X$-rays) is generally interpreted as
being due to synchrotron radiation from a population of non-thermal electrons, whereas the second component
($\gamma$-rays) is explained either through inverse Compton scattering of the same electron
population with the various seed photon fields traversed by the jet~\cite{Bloom}, or by the decay of
neutral pions produced when the highly
relativistic baryonic outflow collides with diffuse gas targets moving across the jet~\cite{Dar:1996qv}.

In this work we focus on the ``relativistic jet meets
target'' scenario, in which $G$-hadrons can be produced in collisions of ultra-high
energy protons in the jet with those in surrounding
gas. In the course of these collisions pions are produced which, on decay,
give rise to a flux of photons and neutrinos. In what follows, we estimate the
relative probabilities for production of $G$'s and high energy neutrinos. We can
then assess whether existing limits on the flux of high energy energy neutrinos and
EGRET data from low energy gamma rays are consistent with the $G$ flux given in
Eq.~(\ref{intflux}).

In order to specify detection criteria, the following kinematic analysis is relevant.
The average energy of the produced $G$ in the target system is given by
\begin{equation}
E_G^{\rm lab} \simeq \sqrt{\frac{E^{\rm lab}_p}{2\ M_p}}\ E_G^{\rm cm} \,\,,
\label{kin1}
\end{equation}
where $E^{\rm lab}_p$ is the energy of the high energy proton in
the jet, and  $E_G^{\rm cm}$ is the $G$ energy in the center-of-mass of the
$pp$ collision. Full acceptance at PAO requires a minimum energy
$E_{G, \,{\rm min}}^{\rm lab}\simeq 5\times 10^9$~GeV for the
showering $G$ hadron. From Eq.~(\ref{kin1}), this determines a
minimum energy for the high energy proton:
\begin{equation}
E^{\rm lab}_p > 2\times 10^{14}\ {\rm GeV}\ \left(\frac{500\ {\rm
GeV}}{M_G}\right)^2 \ \left(\frac{4\ M_G^2}{\hat s}\right),
\label{emin}
\end{equation}
where we have made use of the fact
that, on the average, $E_G^{\rm cm} = \sqrt{\hat s}/2,$ where
$\hat s$ is the square of the energy in the center-of-mass of the
parton-parton collision producing the $G$'s.

It is immediately apparent from Eq.~(\ref{emin}) that ultra-high
energy sources are required in order to produce $G$'s
which can generate extensive air showers. The required energy can
be reduced by restricting production to large $\hat s.$ In order
that the maximum energy at the source does not exceed $10^{14}$ GeV, a
speculative number sometimes used in the literature~\cite{Eberle:2004ua}, we take $\hat
s \ge 16 M_G^2$, which via Eq.~(\ref{emin}) leads to $E^{\rm
lab}_{p, {\rm min}} \approx 5\times 10^{13}\ {\rm GeV}.$

We now turn to evaluating the consistency of the required $G$ flux
and its accompanying pion flux with existing limits on neutrino
and gamma ray fluxes. The required relationship is
\begin{equation}
\int J_{\nu}(E_\nu)\ dE_{\nu} = \frac{2}{3} \,\frac {\sigma_{\rm
inel}}{\sigma_{pp\rightarrow G}(\hat s_{\rm min})}\ \frac{\langle
N_{\nu}\rangle}{N_G}\ \int J_G(E_G)\ dE_G\ \ , \label{jj}
\end{equation}
where $J_{\nu}$ is the neutrino flux, $\sigma_{\rm inel}$ and
$\sigma_{pp\rightarrow G}$ are the total $pp$ inelastic and
inclusive $pp\rightarrow G$ cross sections, and  $N_\nu$ and
$N_G=2$ are the neutrino and gluino multiplicities per collision. The
factor of 2/3 accounts for the fact that only charged pions contribute
to neutrino production. To get our estimates we adopt $\sigma_{\rm inel} \sim 130$~mb~\cite{Alvarez-Muniz:2002ne}.

The  inclusive $pp \to G$ production cross section  has been
evaluated~\cite{Howie} using CTEQ5L parton distribution
functions~\cite{Lai:1999wy}. For $M_G = 500$~GeV, and integrated
over all values of $\hat s,$ the cross section can be conveniently
parametrized as
\begin{equation}
\sigma_{pp\rightarrow G} = 1.17 \times 10^{-43} \,\,
\left(\frac{E_p}{{\rm GeV}}\right)^{1.0565}~{\rm cm^2}\,\, .
\label{sigma}
\end{equation}
When the condition $\hat s\ge 16 M_G^2$ is imposed on the
integration, we have estimated that the cross section given in
Eq.~(\ref{sigma}) is decreased by a factor of 2. The dependence on
$M_G$ can be roughly described by the scaling behavior $e^{-0.007
(M_G - 500~{\rm GeV})}.$

The $E_G$ integration interval in Eq.~(\ref{jj}) is very narrow, as
discussed above. Since we are working  at the very
high energy end of the cosmic ray
spectrum, we will examine the consistency of Eq.~(\ref{jj}) with bounds on the
neutrino flux at the highest energy. Inserting  Eq.~(\ref{sigma}) evaluated at
$\langle E_p\rangle = 7.5\times 10^{13}$ GeV and appropriately reduced to allow for
the $\hat s_{\rm min}$ cut, as well as Eq.~(\ref{intflux}), into
Eq.~(\ref{jj}), we obtain (with $N_G=2$) for the high energy neutrino flux accompanying
$G$-production
\begin{eqnarray}
J_{\nu}(E_\nu)\ \Delta E_{\nu} &\approx& 1200\ \langle
N_{\nu}\rangle \int J_G(E_G)\ dE_G\\
&\approx & \langle N_{\nu}\rangle\, 1.7\times 10^{-18}\,\, {\rm cm}^{-2} \,\,{\rm
s}^{-1}\,\, {\rm sr}^{-1}\,\,
\label{nug}
\end{eqnarray}
for an event rate of 1 yr$^{-1}$. At the end of the spectrum, we may approximately
consider only the most energetic of the secondary $\pi^{\pm},$ carrying  about
8\% of the primary energy~\cite{Knapp:1996fv}. This entails a neutrino multiplicity
$\langle N_{\nu}\rangle =3$/event, each carrying 1/4 of the pion energy. Thus
$\Delta E_\nu\sim \langle E_\nu\rangle \simeq 0.02\ \langle E_p\rangle =
1.5\times 10^{12}$ GeV. The expected neutrino flux (all flavors) at this energy is
then
\begin{equation}
J_\nu(1.5\times 10^{12}\ {\rm GeV}) \approx 4.4\times
10^{-30}\,{\rm GeV}^{-1} {\rm cm}^{-2} \,\,{\rm s}^{-1}\,\, {\rm
sr}^{-1}\,\, . \label{jnupred}
\end{equation}
An upper bound for the neutrino flux in this energy range has been
obtained through the absence of radio signals originating in the
Moon's rim (GLUE~\cite{Gorham:2003da}) or in the Greenland ice
sheet (FORTE satellite~\cite{Lehtinen:2003xv}).  The flux
associated with 1 $G$ event/yr at PAO, given in
Eq.~(\ref{jnupred}), is comfortably below these limits. The RICE
experiment~\cite{Kravchenko:2003tc} (detecting electron
neutrino-induced radio Cerenkov radiation in the polar ice cap)
has reported upper bounds on the $\nu_e+\bar \nu_e$ flux for
energies up to $10^{12}$ GeV. A slight extrapolation of their
result to $1.5\times 10^{12}$ GeV gives a (3-flavor) upper bound
which is a factor of 2.5 lower than the flux in  Eq.~(\ref{jnupred}).
The flux in Eq.~(\ref{jnupred}) is also a factor of 3 lower than the
model-independent bound found in~\cite{Anchordoqui:2002vb} from
the absence of horizontal air showers. Finally, extrapolation of 
the ultra-high energy neutrino intensity given in Eq.~(\ref{jnupred}) down to lower energies, 
assuming $J_\nu(E_\nu) \propto E_\nu^{-2}$ (more on this below), leads to 
a neutrino flux which is in agreement with all upper limits on $J_\nu(E_\nu)$ 
reported by the AMANDA Collaboration~\cite{Ackermann:2005sb}.

There is an additional bound (known as the cascade limit) coming
from EGRET's observation of GeV gamma
rays~\cite{Sreekumar:1997un}. The relation to neutrinos is as
follows: isotopically symmetric triplets of $\pi^+,$ $\pi^-,$ and
$\pi^0$ produced at high energy sources yield 3 $\nu$'s per
charged pion and 2 $\gamma$'s per $\pi^0,$ with energies $E_\nu =
E_\pi/4$ and $E_\gamma = E_\pi/2,$ respectively. ($E_\pi$ is the
pion energy.) The integrated neutrino and gamma ray energies then
satisfy~\cite{Anchordoqui:2004eu}
\begin{equation}
\int_{E_{\pi,{\rm min}}/4}^{E_{\pi,{\rm max}}/4} E_\nu\ J_\nu(E_\nu)\,\, dE_\nu = \frac{3}{2}
\int_{E_{\pi,{\rm min}}/2}^{E_{\pi,{\rm max}}/2} E_\gamma\ J_\gamma(E_\gamma)\,\, dE_\gamma \ \ .
\label{eflux}
\end{equation}
Normalization to EGRET data~\cite{Sreekumar:1997un},
\begin{equation}
\int_{0.1\ {\rm GeV}}^\infty J_\gamma(E')\,\, dE' = 1.45\times 10^{-5}\,\, 
{\rm cm}^{-2} \,\,{\rm s}^{-1}\,\, {\rm
sr}^{-1}\,\, ,
\end{equation}
with a spectrum $J_\gamma(E_\gamma) = C_\gamma E_\gamma^{-2}$,
implies $C_\gamma = 1.45\times 10^{-6}\,{\rm GeV}\ {\rm cm}^{-2}
\,\,{\rm s}^{-1}\,\, {\rm sr}^{-1}.$ Using Eq.~(\ref{eflux}) we
obtain $C_\nu = 3/2 C_\gamma$ for the normalization constant of
the accompanying neutrino spectrum, again on the assumption of a
spectrum  $J_\nu\propto E_\nu^{-2}.$ At face value this gives a
flux at $1.5\times 10^{12}$ GeV which is nearly a factor of 5
smaller than the flux in Eq.~(\ref{jnupred}). However, this  
represents a  very large extrapolation: even logarithmic corrections 
to the $E^{-2}$
spectrum   could result in sizeable deviations at very high
energies.  To see how such corrections might arise, we note that
for a primary high energy proton with energy $E_p$, the resulting
pion spectrum is expected to obey a modified Feynman scaling in
the central rapidity region, $dN_\pi/dE_\pi|_{E_p} \approx
C(E_p)/E_\pi,$ where $C$ may be growing as some power of
$\ln(E_p)$~\cite{Abe}. For given $E_\pi\ < 0.08 \,E_{p,{\rm
max}},$ we may convolve with a proton spectrum typical of Fermi
engines, $ dN_p/dE_p \propto 1/E_p^2,$ to obtain the pion
spectrum~\cite{nuX}:
\begin{equation}
\frac{dN_\pi}{dE_\pi}  =  \int_{E_\pi/0.08}^{E_{p, {\rm max}}}
dE_p\; \left. \frac{dN_\pi}{dE_\pi}
\right|_{E_p}\frac{dN_p}{dE_p}\;\; \propto\;\;  \frac{\bar
C(E_\pi)}{E_\pi^2}\,\, , \label{soft1}
\end{equation}
where $\bar C(E_\pi)$ is generically a function which grows as a
power of $\ln(E_\pi)$, falling to zero at the cutoff $E_\pi=0.08\
E_{p,{\rm max}}$. Hence we suspend judgement with respect to the
constraint imposed by the cascade bound.

\section{$\bm{\Lambda_{\rm SUSY}}$ written in the sky?}
\label{lambdaS}

We have described conditions under which $G$-hadrons could be produced in
astrophysical sources in sufficient abundance to allow detection in an
air shower array. Moreover, the complete longitudinal and shower-front
curvature profiles of $G$-hadron cascades would uniquely differentiate them
from proton and nucleus backgrounds. In this last part of the paper we examine
what we can learn about SUSY if such events are actually observed.

We have seen that limits on heavy isotope abundance place an {\em
upper} bound on $\Lambda_{\rm SUSY}$. Detection of $G$-hadrons,
presumed to originate at cosmological distances, will place a
{\em lower} bound on the proper lifetime $\tau_0$ of the $G$.
With
the use of Eq.~(\ref{tau0}) this can be translated into a lower bound
on the scale of SUSY breaking. The argument can be specified as
follows. In the presence of decay, the integration of the $G$-flux over source distances
$r$ out to the horizon ${\cal R}\approx 3$ Gpc is modified by inclusion of a damping
factor
\begin{eqnarray}
f&=&\int_0^{\cal R} e^{-r/[c\tau(E_G)]} dr\\
&=& \frac{c\tau(E_G)}{{\cal R}}\,\left(1-e^{-{\cal R}/[c\tau(E_G)]}\right)\ \ ,
\end{eqnarray}
where the Lorentz dilated $G$ lifetime is given in terms of the proper lifetime
$\tau(E)=\tau_0\,E_G/M_{G},$ with
$E_G\approx 5\times 10^9\ {\rm GeV}.$ If a few $G$ events are seen during the
lifetime of PAO,  and the observation of $>10^{12}$ GeV neutrinos  merits the expectation of
1 $G$ event/yr, then $f$ cannot be too small, say $f\agt 0.1.$ This
implies that $c\tau(E_G)/{\cal R} \agt 0.1,$ and places a bound
\begin{equation}
\tau_0 \agt 100\ {\rm yr}\ \
\end{equation}
on the proper lifetime. From Eq.(\ref{tau0}), one
would then obtain a {\em lower} limit on the SUSY breaking scale,
\begin{equation}
\Lambda_{\rm SUSY} > 2.6\times 10^{11}\ {\rm GeV}\ \ .
\end{equation}
In conjunction with the upper limit from isotope abundance,
$\Lambda_{\rm SUSY} < 10^{13}\ {\rm GeV},$ we can then adduce the remarkable
result that {\em the observation of a few $G$ events during the operating
life of PAO can fix the scale of high energy SUSY breaking.}

\section{Summary}
\label{conclusions}

In this work we have examined under which conditions metastable
$G$-hadrons can be synthesized in powerful astrophysical
environments and can be detected on Earth. We show, that if cosmic 
ray sources are able to accelerate protons somewhat above 
$5 \times 10^{13}$~GeV, about 0.5-1 $G$-hadron induced
shower per year could be detected at PAO. Additionally, $G$-cascades
provide a signal which is easily differentiated from background. These
numbers are for $M_G=500$ GeV: the event rate could be doubled for
$M_G=400$ GeV.

Should some of these very distinctive showers be
observed, a {\it lower} bound on $\Lambda_{\rm SUSY}$ can be
deduced in a manner unavailable at colliders. The combination of
such a lower bound  with the upper bound imposed by the scarcity
of heavy isotopes fixes the scale of SUSY breaking to a
relatively narrow window, $2.6 \times 10^{11}~{\rm GeV} < \Lambda_{\rm SUSY}
< 10^{13}~{\rm GeV}.$ This result, if true, combined with the
expected low mass of the higgs would provide strong support for a
finely-tuned universe.\\

\underline{Note added:} After this paper was completed, a work
with related interesting ideas for detection of long-lived
gluinos at colliders and at IceCube appeared~\cite{Hewett:2004nw}.
For observation at IceCube, the gluino production takes place in
cosmic ray collisions in the Earth's atmosphere. The steeply falling
cosmic ray luminosity above $10^9$~GeV allows sufficient $G$ production
for a measurable signal at IceCube only for $M_G\alt$ 150 GeV,
which is very close to the
present  bounds from Tevatron~\cite{Acosta:2002ju}. In examining
production at extraterrestrial sources, the present work can substantially
extend the possible range of $M_G$ able to be probed by cosmic ray
experiments.

\begin{acknowledgments}

We are grateful to Howie Baer for  valuable discussion and for
providing cross section estimates for high energy $G$-production.
We have also benefited from
discussions with Concha Gonzalez-Garcia and
Darien Wood. The work of LAA and HG has been partially supported by the
US National Science Foundation (NSF), under grants No.\
PHY--0140407 and No.\ PHY--0244507, respectively. The work of CN
is supported by a Pappalardo Fellowship and in part by funds
provided by the U.S.Department of Energy (DoE) under cooperative
research agreement DF-FC02-94ER408818.

\end{acknowledgments}



\begin{thebibliography}{99}
\bibitem{nuX} Note that the neutrino spectrum will conform to this as well, i.e.,
$dN_\nu/dE_\nu \propto E_\nu^{-2}.$
\end{thebibliography}
\end{document}